\documentclass[11pt,fleqn]{article}

\usepackage{amsmath} 
\usepackage{amsfonts}
\usepackage{graphicx}
\usepackage{dcolumn}
\usepackage{upgreek}
\usepackage{latexsym,slashed}
\usepackage{amssymb,amsfonts}
\usepackage{cancel}
\usepackage{bm} 
\usepackage{amsmath,mathrsfs} 

\newcommand{\address}[1]{\begin{center}\large #1\end{center}}

\def\beq{\begin{eqnarray}}
\def\eeq{\end{eqnarray}}

\def\R{{\hbox{{\rm I}\kern-.2em\hbox{\rm R}}}} 
\def\H{{\hbox{{\rm I}\kern-.2em\hbox{\rm H}}}} 
\def\N{{\hbox{{\rm I}\kern-.2em\hbox{\rm N}}}} 
\def\C{{\ \hbox{{\rm I}\kern-.6em\hbox{\bf C}}}} 
\def\Z{{\hbox{{\rm Z}\kern-.4em\hbox{\rm Z}}}} 
 
\catcode`\@=11
\@addtoreset{equation}{section}

\begin{document}
\tolerance=5000

\title{\bf{ Ray tracing in FLRW flat space-times}} 

\author{
Giovanni~Acquaviva$\,^{(a)}$\footnote{gioacqua@gmail.com},
Luca~Bonetti$\,^{(a)}$\footnote{bonetti659@gmail.com},
Guido~Cognola$\,^{(a)}$\footnote{cognola@science.unitn.it}, and
Sergio~Zerbini$\,^{(a)}$\footnote{zerbini@science.unitn.it}}
\date{}
\maketitle

\address{$^{(a)}$ Dipartimento di Fisica, Universit\`a di Trento \\
and Istituto Nazionale di Fisica Nucleare - Gruppo Collegato di Trento\\
Via Sommarive 14, 38123 Povo, Italia}
\medskip 
\medskip

\begin{abstract}
 In this work we take moves from the debate triggered by Melia \textit{et al.} in \cite{melia} and followed by opposite comments by Lewis and Oirschot in \cite{lewis,lewis1}.  The point in question regards the role of the Hubble horizon as a limit for observability in a cosmological setting.  
We propose to tackle the issue in a broader way by relating it to the causal character of the Hubble surface and to the tracing of null trajectories, focusing on both three-fluids and generalized Chaplygin gas models.  The results should make clear that for quite reasonable and physically motivated models, light rays reaching a comoving observer at $R(t_0)=0$ have never traveled a distance greater than the proper radius of the horizon until $t_0$.
\end{abstract}

\section{Introduction}\label{intro}

Relativistic theories of gravity  on flat FLRW space-times have become important in modern cosmology after the discovery of the current cosmic acceleration, the rising of the dark energy issue and the confirmation of inflationary models. Among the several descriptions of the current accelerated expansion of the universe, the simplest one considers the introduction of a small positive cosmological constant in the framework of General Relativity, so that one is dealing with a perfect fluid whose equation of state parameter $\omega=-1$. This fluid model is able to describe the current cosmic acceleration.  Also other forms of fluid (phantom, quintessence, inhomogeneous fluids, etc.) satisfying suitable equation of state are not excluded, since the observed small value of cosmological constant leads to several conceptual problems -- the debate on vacuum energy and the coincidence problem, among others.
For this reason, several different approaches to the dark energy issue have been proposed. 
Among them, modified theories of gravity~\cite{Review1}--\cite{rev} represent an interesting extension of Einstein's theory.  Unfortunately, a large class of these modified models admit future singularities, the worst being the so-called Big Rip singularity \cite{brip} (for a general discussion, see for example \cite{tsu}) .

With these models in mind, we revisit in a deep and analytic way the analysis -- proposed first in \cite{ellis} and recently reproposed in the context of the debate \cite{melia,lewis,lewis1} -- of light trajectories in FLRW models and the role of the Hubble horizon as an observational limit for comoving observers, hopefully elucidating some points.\\
We restrict our analysis to a flat FLRW model, which is also a spherically symmetric dynamical space-time admitting a dynamical horizon.  
For the sake of completeness, we briefly review the general formalism \cite{kodama,hayward,sean09,bob09,Vanzotop} that will be useful in the following. 

Recall that any  spherically symmetric dynamical space-time has a metric which can locally be expressed in the form
\beq
\label{metric}
ds^2 =\gamma_{ij}(x)dx^idx^j+ R^2(x) d\Omega^2\,,
         \qquad i,j=0,1\,,\qquad x=\{x^i\}\equiv\{x^0,x^1\}\;,
\eeq
where  the two-dimensional metric
\beq d\gamma^2=\gamma_{ij}(x)dx^idx^j
\label{nm}
\eeq
is referred to as the ``normal'' metric, 
$\{x^i\}$ being the coordinates of the corresponding two dimensional ``normal'' space 
and $R(x)$ the areal radius, which is
a scalar quantity in the normal space. Finally $d\Omega^2$ is the metric of a
two-dimensional sphere $S_2$. Associated with the areal radius, there exists a spherical surface $S(x)=4\pi R^2(x)$. 
It will be useful to define also the expansions related to the horizon  surface, that is the rate of change of the area transverse to bundles of null rays orthogonal to the horizon.  In spherical symmetry and  double null coordinates,  the two expansions  are given by
\begin{equation*}
 \theta_{\pm} = \frac{\partial_{\pm} S}{S}=\frac{2}{R}\, \partial_{\pm} R\, ,
\end{equation*}
where $R$ is the areal radius.  A marginal surface is defined by
$\theta_+=0$, and it is \textit{future} if $\theta_-<0$ and
\textit{past} if $\theta_->0$.  
Moreover the sign of $\partial_-\theta_+$ discerns whether the horizon is \textit{inner} (positive) or \textit{outer} (negative).

To make use of a covariant formulation,  one may introduce the  normal space scalar quantity proportional to $\theta_+\theta_-$, namely
\beq
\Phi(x)=\gamma^{ij}(x)\partial_i R(x)\partial_j R(x)\,, \label{sh}\,. 
\eeq 
The surface $S$ is said trapped if the related scalar $\Phi (x)<0$, untrapped  if $\Phi (x)>0 $, and marginal if  $\Phi (x)=0$.   
The dynamical trapping horizon according to Hayward  is a surface foliated by the marginal surfaces, namely   
is  the solution of  equation
\beq 
\Phi(x)\Big\vert_{x=x_H} = 0\,, \label{ho} 
\eeq
provided that $\partial_i\Phi\vert_{x=x_H}\neq 0$. 
The trapping horizon is a quite natural generalization of the event horizon, which, in the dynamical setting,  is ``teleological'' in its 
definition \cite{abby}. 

For the sake of clarity, we give two examples. The first one is the static metric describing  Schwarzschild space-time 
\beq
ds^2=-\left(1-\frac{2M}{r}\right)dt^2+\frac{dr^2}{(1-\frac{2M}{r})}+  r^2\,d\Omega^2\,. 
\eeq
In this gauge the coordinates are $x=(t,r)$, the areal radius concides with $r$ and the normal metric
\beq
d\gamma^2=-\left(1-\frac{2M}{r}\right)dt^2+\frac{dr^2}{(1-\frac{2M}{r})}\,. 
\eeq
The horizon is a static one and, in this case, concides with the event horizon: it is given by equation (\ref{ho})
\beq
\Phi\vert_H=1-\frac{2M}{r_H}=0\,, 
\eeq
namely one gets the usual Schwarzschild radius  $r_H=2M$. We also stress the fact that the formalism is covariant. For example in the Painleve's system of coordinates
$(v, r)$, with a different time coordinate $v$, the normal metric is static but not diagonal, namely
\beq
d\gamma^2=-\left(1-\frac{2M}{r}\right)dv^2-2\sqrt{\frac{2M}{r}} dv dr\,. 
\eeq
Again, the horizon is located at $r_H=2M$, but now the normal metric evaluated on the horizon is regular and null.

The second example is the one we are mainly interested in. Let us consider the  flat FLRW space-time,  the metric usually being written in the form 
\beq
ds^2=-dt^2+a^2(t)\left(dr^2+r^2\,d\Omega^2 \right)\,.
\eeq
The coordinates are $x=(t, r)$, the areal radius is $R=a(t)\,r$ and the normal metric simply reads 
\beq
d\gamma^2=-dt^2+a^2(t)dr^2\,.
\eeq
Thus, 
\beq
\Phi\vert_{H}=\left[ -(\partial_tR)^2+\frac{1}{a^2(t)}(\partial_rR)^2 \right]_{H}=-\dot a^2r^2+1\Big\vert_H=0\,,\quad\quad \dot a=\frac{da}{dt}\,,
\eeq
namely the trapping horizon is located at $r_H=1/\dot{a}$, and in terms of areal (or proper) radius reads
\beq
R_H = a(t)\,r_H=\frac{1}{H(t)}\,,
\eeq
where the Hubble parameter $H(t)$ is defined by
\beq
H(t)=\frac{\dot a}{a}=\frac{d \ln a}{dt}\,.
\label{h}
\eeq
The quantity $R_H$ is known as the Hubble sphere, but we may also refer to it as the  Hubble dynamical horizon in the Hayward terminology. 
In this true dynamical case, the normal metric evaluated on the dynamical horizon reads
\beq
d\gamma^2_H=-dt^2+a(t)^2(dr_H)^2\frac{\dot{H}^2+2\dot{H}
  H^2}{H^4}dt^2=
\left[\left(\frac{d R_H}{dt}\right)^2-2\frac{d R_H}{dt} \right]dt^2\,.
\label{coglio}
\eeq
The sign of the line element, related to the value of  the quantity $\dot R_H$,
determines the causal character of the horizon surface, which is 
timelike, spacelike or null according to whether  $d\gamma_H^2$
is negative, positive or vanishing respectively.
For example,  in the de Sitter case, $R_H=1/H_0$, and thus the
correponding horizon is null, as for a generic static black hole. As already stressed by 
other authors, when $d\gamma_H^2\neq0$,  photons may cross the dynamical horizon several times, but this is not surprising and 
this property depends on the cosmological model 
(see the  examples presented in Sec. (\ref{models}), and in \cite{lewis1}). 
The horizon surface is always space-like for decreasing $R_H$. This case will be discussed in detail in Section  \ref{causal}  of the paper.  

The paper is organized as follows. In Sec. \ref{models} we review
some past and future singularity scenarios in a flat FLRW universe.  
In Sec. \ref{causal} we analyze the causal character of cosmological horizons: this in turn introduces the topic of Sec. \ref{rays}, where the ray tracing of null trajectories is discussed.  Conclusions are given in Sec. \ref{conc}.

\section{From $\Lambda$CDM to  Big Rip solutions}
\label{models}

Here we review the conditions under which cosmological past and future singular solutions like Big Bang, Little Rip and Big Rip may be present.  We recall the form of flat FLRW space-time in the spherical coordinates $(t, r, \theta , \psi) $
\begin{equation}\label{frw}
 ds^2 = -dt^2 + a^2(t) \left( dr^2 + r^2 d\Omega^2 \right)=d\gamma^2+a^2(t)r^2 d\Omega^2 \,.
\end{equation}
For our discussion it is convenient to introduce the coordinate defined by the proper radius $R=r\, a(t)$. Thus, one has
\beq
ds^2 = -(1-H^2R^2)dt^2 -2RHdRdt+dR^2 + R^2 d\Omega^2\,.
\eeq
This expression suggests the introduction as evolution parameter of the quantity  $y=\ln a(t)$,  largely used  in inflationary and dark energy  models (for example, see the recent paper \cite{guo}) . As a result, the cosmic time may be expressed as
\beq
t(y)=\int \frac{dy}{H(y)}\,
\label{t}
\eeq
and the normal metric, the only relevant for our discussion, in the new coordinates $(y, R)$ reads
\beq
d\gamma^2=-\left(\frac{1}{H^2}-R^2 \right)dy^2-2Rdy dR+dR^2\,.
\label{y}
\eeq
It is easy to check that the trapping horizon is again given by the Hubble horizon $R_H=\frac{1}{H}$, and we may rewrite
\beq
t(y)=\int dy R_H(y)\,.
\label{tt}
\eeq
We must supply this ``new kinematic'' FLRW  framework with the dynamics of gravity. 
 
One may  assume a generalization of the  Friedmann equation  and matter energy conservation together with a suitable equation of state, namely
\begin{align}
 &H^2=\frac{\chi}{3} F(\rho)\,,\\
 &\frac{d\rho}{dy}+3(p+\rho)=0\,, \quad  p=p(\rho)\,.
\label{mc}
\end{align}
Here $\chi=8\pi G$. In general $F(\rho)$ has  to be non-negative. For further generalizations of  Friedmann equation see \cite{cognola13} and references therein.

\subsection{Standard equation of state} 
Recall that in general relativity  $F(\rho)=\sum_i \rho_i$ is  linear
in the density species.  
With $\omega_i $ constant quantities one has the simple equations of state 
\beq
p_i=\omega_i \rho_i\,,
\eeq
and assuming matter conservation for every species one gets
\beq
\frac{d \rho_i}{d y}=-3(1+\omega_i) \rho_i \,,
\eeq 
which can be solved to give
\beq
\rho_i=c_i e^{-3(1+\omega_i)y}\,.
\eeq 
We thus have
\beq
H^2=\frac{\chi}{3}\sum_i c_i e^{-3(1+\omega_i)y}\,.
\eeq
The associated Hubble horizon is 
\beq
R_H(y)=\sqrt{\frac{3}{\chi}}\frac{e^{3y/2}}{(\sum_ic_i e^{-3\omega_i y})^{1/2}}\,.
\label{rh1}
\eeq
As a consequence, the solution of Friedmann equation may be expressed as
\beq
t(y)=t(y_0)+\sqrt{\frac{3}{\chi}}\int_{y_0}^y dx \frac{e^{3x/2}}{(\sum_ic_i e^{-3\omega_i x})^{1/2}}\,.
\label{fs}
\eeq
\paragraph{One-fluid model.}

The simplest example is the one-fluid model with equation of state
$p=\omega\rho$. In such a case  one has 
\beq
t(y)=t(y_0)+\sqrt{\frac{3}{c\, \chi}}\int_{y_0}^y dx\, e^{3(1+\omega)x/2}\,.
\label{fs0}
\eeq
If $1+\omega >0$, then we may choose $y_0=-\infty$ with $t(-\infty)=0$ (the Big Bang) and so
\beq
t(y)=\sqrt{\frac{3}{c\, \chi}}\,
\frac{2}{3(1+\omega)}\,\,e^{3(1+\omega)y/2}=\frac{2}{3(1+\omega)}\,\,a^{3(1+\omega)}\,, 
\eeq
which, after inversion, gives the usual flat FLRW solution as a function of the time $t$. 
In the special case $\omega=-1$ there is no Big Bang and from (\ref{fs0})
one trivially gets the de Sitter solution $t \simeq y= \ln a$.

\paragraph{Three-fluids model.} Now we consider an interesting
phenomenological generalization of previous case
describing (dark) matter, radiation  and dark energy (cosmological
constant or phantom matter). 
 The total energy density and equations of state read
\begin{equation}\label{doublefluid}
\rho_T=\rho_m+\rho_r+\rho_f\,, \quad p_m=0\,, \quad  p_r=\frac13\rho\,, \quad p_f=\omega_f \rho \,. 
\end{equation}
 For phantom matter, we make the choice  $1+\omega_f=-\delta <0$. The energy-matter conservation gives
\begin{equation}
\rho_m=c_0\, e^{-3 y}\,, \quad \rho_r=c_r\, e^{-y}\,, \quad \rho_f=c_f\, e^{-3(1+\omega_f)y} \,.
\end{equation}
From the Friedmann equation we have 
\begin{equation}
R_H(y)= \sqrt{\frac{3}{\chi}} \frac{e^{3y/2}}{\left(c_0+c_r e^{-y}+ c_f e^{(3+3\delta) y} \right)^{1/2}} \,
\label{mph}
\end{equation}
and hence
\beq
\sqrt{\frac{\chi}{3}}\,t(y)=\int_{-\infty}^y  \frac{e^{3x/2}\,dx}{\left(c_0+c_r e^{-x}+ c_f e^{(3+3\delta) x} \right)^{1/2}}\,, 
\label{3m}
\eeq
where $y_0=-\infty$, and $t(-\infty)=0$  because here  the integrand is summable. This is the initial Big Bang singularity of this model. 

On the other hand,  the behaviour for large $y$ characterizes the future singularities. In fact, for $\delta=0 $ (the so called $\Lambda$CDM model), $t \rightarrow \infty$ as soon as $ y \rightarrow \infty $. 

In the case of phantom component, $\delta >0$ and small, the integral converges for  $y \rightarrow \infty $. As a result a singularity is present for $y$ and $a$ at a future finite time given by
\beq
\sqrt{\frac{\chi}{3}}\,t_s=\int_{-\infty}^{\infty} \frac{e^{3/2 x}}{\left(c_0+c_r e^{-x}+ c_f e^{(3+3\delta) x} \right)^{1/2}}dx\,. 
\label{3br}
\eeq
This is the well known  Big Rip singularity associated with the presence of a phantom fluid \cite{brip}. In a two-fluids model, namely putting 
$c_r=0$, one has $c_f=1-c_0$,  and $H^2_0=\frac{\chi}{3}$ (the Hubble parameter is a constant). 
The integral can be computed and reads
\beq
t_s=\frac{1}{\sqrt{\pi}H_0}\frac{\Gamma\left(\frac{\delta}{2(1+\delta)}\right)}{(1+\delta) c_0^\delta}\Gamma\left(\frac{1}{2(1+\delta)}\right)
(c_0\,c_f)^{-\frac{1}{2(1+\delta)}}\,. 
\eeq
For small value of $\delta$ one easily gets
\beq
t_s\simeq \frac{1}{H_0}\Big[c_0(1-c_0)\Big]^{-1/2}\Gamma\left(\frac{\delta}{2}\right)\,.
\eeq
Thus, the smaller is $\delta$, and in the future the finite singularity will be located with respect to $\frac{1}{H_0}$ (roughly the age of our universe).

Coming back to (\ref{3m}), its inversion would give the FLRW solution for the $\Lambda$CDM model. As  is well known,  in general the inversion of this equation is a difficult task and numerical analysis is required. 
In the next Section we shalll see that the inversion will not be strictly necessary.

However in a two-fluids model (matter or radiation, plus cosmological constant $\delta=0$),  the inversion is possible since one has (here $\omega$ is either $0$ or $1/3$)
\beq
\sqrt{\frac{\chi}{3}}\,t(y)=\int_{-\infty}^y  \frac{e^{3x/2}}{\left(c\, e^{-3\omega x}+ c_f\, e^{3 x} \right)^{1/2}}dx\, =\, \frac{2}{3(1+\omega)\sqrt{c_f}} \sinh^{-1}\left(e^{3(1+\omega)y/2} \right)\,, 
\label{2m}
\eeq
and this gives the well known result
\begin{equation}
 a(t)=\left(\frac{c}{c_f}\right)^{\frac{1}{3 (1+\omega )}} \sinh\left(\sqrt{3\chi \frac{c}{c_f}}\, \frac{(1+\omega)}{2}\, t\right)^{\frac{2}{3 (1+\omega )}} \,. \label{1}
\end{equation}

\subsection{Modified equation of state}
Another possibility that has been investigated by several authors  is to keep the Friedmann equation with matter conservation but to modify the equation of state, for example considering  
\begin{equation}
p=\omega \rho-A \rho^{-\gamma}\,,  \quad A>0\,.\label{modeos}
\end{equation}
As a result one has
\beq
-3 y= \int \frac{\rho^\gamma}{(1+\omega)\rho^{\gamma+1}-A} d\rho \,.
\label{cp}
\eeq\\

Let us consider first a generalized model for dark energy, where $\omega=-1$ (see, for example, \cite{brevik})
Considering $\gamma=-b-\frac{1}{2}$, one has
\beq
y=\frac{2}{3A(1-2b)}\rho^{1/2-b}\,, \qquad  \rho=\left(C_b\, y\right)^{\frac{2}{1-2b}}\,, 
\eeq
where $C_b=3A(1-2b)/2$.  For $b=0$, one has the so-called Little Rip behaviour $R_H=\sqrt{\frac{3}{\chi}}\frac{3A}{2y}$ (see \cite{frampton}), there is no Big Bang and it is possible to show that
\begin{equation}
H(t)=H_0\, e^{B(0) t}\,,
\end{equation}
with $B$ constant.
If $b \neq 0$, the solution may be written in the form
\begin{equation}
H(t)=H_0 \Big(1-2b B(b)(t-t_0) \Big)^{-\frac{1}{2b}}\,.
\end{equation}
If $b<0$ one has a Little Rip  singularity, but if $b>0$ one has a Big Rip singularity \cite{brevik}.\\

If $\omega+1>0$, then one is dealing with a Chaplygin gas and its generalizations \cite{ugo,bento}. In this case, one obtains
\beq
\rho=e^{-3y}\left(\frac{1+Ae^{3\alpha y}}{(1+\omega)}  \right)^{\frac{1}{\alpha}}\,, \quad \alpha=(1+\omega)(1/2-b)>0\,.
\label{ch}
\eeq
The Hubble horizon is $R_H=\frac{1}{H}$, and the time reads
\beq
\sqrt{\frac{\chi}{3}}t(y)= \int_{-\infty}^y \,dx e^{3/2x} \left(\frac{1+Ae^{3\alpha x}}{(1+\omega)}  \right)^{-\frac{1}{2\alpha}}\,.
\label{tch}
\eeq
There is a Big Bang singularity, but no future singularity because the above integral diverges as $y \rightarrow \infty$.

\section{Ray tracing in FLRW space-times}
\label{rays}
We here describe the null trajectories followed by light rays in flat FLRW.  This analysis allows to visualize the range of possible trajectories followed by massless  bodies in a given space-time.  In particular it is possible to determine i) whether or not a comoving observer sitting in the origin at time $t_0$ will receive ingoing light rays and ii) the maximum proper radius reached by these light rays before $t_0$.  In the following, we reformulate the analysis presented in \cite{ellis} and recently in \cite{melia,lewis}, with the main aim to present analytic results.\\

From equation (\ref{y}) we find for radial ingoing photon geodesics 
\begin{equation}\label{ode}
\frac{d R_{\gamma}}{dy} =R_{\gamma}- \frac{1}{H}\equiv R_{\gamma}-R_{H} \,.
\end{equation}
  The general solution of eq.(\ref{ode}) is given by
\begin{equation}
 R_{\gamma}(y) =e^{y} \left( C- \int^y_{-\infty}e^{-x} R_H(x)\, dx \right)\,.
\label{rt}
\end{equation}
Here we have assumed the existence of a Big Bang initial singularity $y_0\rightarrow-\infty$.
Providing the model through the specification of $R_H$ and appropriate initial conditions, one can trace ingoing light rays.\\

First we  discuss  the Hubble horizon behavior. In the standard one-fluid model, one has 
\beq
H^2=\frac{\chi}{3} \rho=H^2_0 e^{-3(1+\omega)y}\,, \quad H^2_0=\frac{\chi c}{3}\,.
\eeq
Thus, the Hubble horizon is always expanding according to
\beq
R_H=\frac{1}{H_0} \,e^{3(1+\omega) y/2}\,.
\eeq
For the generalized Chaplygin case we have an increasing  but asymptotically constant function  in $y$
\beq
R_H(y)=\sqrt{\frac{3 (1+\omega)^{1/\alpha}}{\chi}}\frac{e^{3y/2}}{\left(1+A\,e^{3\alpha y}\right)^{\frac{1}{2\alpha}}}\,.
\label{rhc}
\eeq
Here $R_H(-\infty)=0$, which is the Big-Bang singularity.  For $y \rightarrow \infty$,  $R_H(y)$  reaches its maximum given by
\beq
R_H^{max}=  \sqrt{\frac{3 (\frac{1+\omega}{A})^{1/\alpha}}{\chi}}\,.
\eeq
A similar behaviour is present for the three-fluids model in the case $\delta=0$, see (\ref{mph}), and the maximum for $y\rightarrow\infty$ now reads
\beq
R_H^{max}=  \sqrt{\frac{3}{\chi\,  c_f}}\,.
\eeq
In the case of phantom field ($\delta >0$), since for $y \rightarrow \infty$  one has $R_H(y) \rightarrow 0 $, it follows that there exists a local maximum  at finite $y=y^*$, 
given by the solution of the transcendental equation
\beq
3\, c_0 + 4\, c_r\, e^{-y^*}= 3\, \delta\, c_f\, e^{(3+3\delta)y^*}\, .
\label{max}
\eeq
At the Big Bang $y\to-\infty$ one has
\beq
\frac{d R_H}{dy}\Big\vert_{y=-\infty}=0\,, \qquad \frac{e^{-y}d R_H}{dy}\Big\vert_{y=-\infty}=0\,.
\eeq\newline

With regard to photon  tracing, in the standard one fluid model \cite{ellis} 
\begin{equation}
 R_{\gamma}(y) =e^{y} \left( C- \frac{2}{(1+3\omega)H_0}e^{(3\omega+1)y/2} \, \right)\,.
\label{rtsm}
\end{equation}
The photon trajectory, chosen an arbitrary $C$,  always reaches the origin again, namely  $R_\gamma(y_1)=0$ at 
\beq
H_0C=e^{(3\omega+1)y_1/2}
\eeq
In presence of dark energy, the situation changes. In fact, in the three-fluids and generalized Chaplygin models,
 having Big Bang singularities, one has 
\beq
\frac{d R_\gamma}{dy}\Big\vert_{y=-\infty}=0\,, \quad \frac{e^{-y}d R_\gamma}{dy}\Big\vert_{y=-\infty}=C\,.
\eeq
which gives a physical meaning to the integration constant $C$. Furthermore, in these cases,  the crucial fact is the existence of the finite integral 
\beq
C^*=\int^{\infty}_{-\infty}e^{-x} R_H(x)\, dx\, < \infty\,.
\label{C}
\eeq
The corresponding constant in the one-fluid model is obviously divergent.   For the two fluids model one has
\beq
C^*(\delta)=\frac{1}{\sqrt{\pi}H_0}\frac{\Gamma\left(\frac{2+3\delta}{6(1+\delta)}\right)}{6(1+\delta)}\Gamma\left(\frac{1}{6(1+\delta)}\right)
\left(\frac{c_0}{c_f}\right)^{\frac{1}{6(1+\delta)}}\,, 
\eeq 
while for the Chaplygin gas
\beq
C^*(\alpha)=R_H(\alpha)\vert_M\, \frac{\Gamma\left(\frac{1}{6\alpha}\right)A^{-1/6\alpha}}{6\, \alpha\, \Gamma\left(\frac{1}{2\alpha}\right)}\Gamma\left(\frac{1}{3\alpha}\right)
\,. 
\eeq 
As a consequence, one can distinguish between three cases.

The first one is the most interesting from the physical point of view and it is realized when $C <C^*$. In this case, for $y \rightarrow  \infty$ 
one has $R_\gamma(y) \rightarrow -\infty $. Of course, only positive values of $R_\gamma$ are physically relevant, thus there exists $y_1$ such that 
\beq
C=\int^{y_1}_{-\infty}e^{-x} R_H(x)\, dx\, ,\quad R_\gamma(y_1)=0\,,
\eeq
 namely these photons emitted at the Big Bang may be observed at the origin after a finite ``time'' $y_1$ and their trajectories are hence given by
\beq
 R_{\gamma} (y) = e^{y}\, \int_y^{y_1} e^{-x}\, R_H(x)\, dx
\eeq 
For this class of trajectories  there exists an extremal $\frac{d R_{\gamma}}{dy}=0$ at $y_M$,
which defines the horizon crossing 
\beq
R_\gamma\vert_M=R_H\vert_M\,.
\eeq
Making use of photon trajectory equation one has on the extremal
\beq
\frac{d^2 R_\gamma}{d^2y}\Big\vert_M=-\frac{dR_H}{d y}\Big\vert_M\,.
\label{ex}
\eeq
In order for the light ray to eventually reach the origin, the moment of the horizon crossing should correspond to a maximum of the trajectory.  From the last equation, this means that $dR_H/dy >0$ at $y_M$, \textit{i.e.} the horizon's proper radius has to be an increasing function in a neighborhood of $y_M$.\\
In both the Chaplygin gas and the $\delta=0$ three-fluids models, the horizon radius is always an increasing function of $y$.  On the other hand, for $\delta>0$ and smaller it has to be $y_M < y^*$ ($y^*$ being the time corresponding to the maximum value of horizon radius $R_H$) because in this range $R_H$ is increasing. 
In general, for this class of photon trajectories one has the trivial but important property (see \cite{melia})
\beq\label{bound}
R_\gamma(y_M)=R_H(y_M) < R_H(y^*)\,.
\eeq
 This property supports the claim put forward graphically in Ref. \cite{melia} and gives an important global geometric characterization of the Hayward trapping horizon $R_H=\frac{1}{H}$.\\

In the other two cases ($C >C^*$ and $C=C^*$) $R_\gamma$ is never vanishing. Furthermore, when  $y \rightarrow \infty$, for $C >C^*$ it follows 
  $R_\gamma(y) \rightarrow \infty$, while for $C =C^*$ one has $R_\gamma(y) \rightarrow 0$.

For the $\delta=0$ and the Chaplygin gas models there are no extremal points. In fact, if there were an extremal, due to equation (\ref{ex}) this should be a local maximum, and that would contradict $R_\gamma(\infty)= \infty$. 

In the phantom case there exists a  first extremal (and we have seen that this is a local maximum), but there exists also a second extremal, which has to be a local minimum in order to be compatible with  $R_\gamma \rightarrow \infty$. 
In any case, this class of  photon trajectories cannot ever be observed at the origin.

\section{Hubble horizon and its causal character}
\label{causal}
The causal characterization of the Hubble horizon in different models can be useful in order to better clarify the behavior of light trajectories -- a topic that we have addressed in previous sections.  In a flat FLRW model the horizon is a spherically symmetric surface located at $r_H\, \dot{a}(t)=1$.  Evaluating the normal metric on the horizon one may rewrite  
\beq
d\gamma^2_H=\frac{d R_H}{dy}\left( \frac{d R_H}{dy}-2 R_H  \right)dy^2\,.
\label{gh}
\eeq
 
Recall that the sign of the line element determines the causal character of the horizon surface, in particular for $d\gamma_H^2<0\, (>0)$ the horizon will be timelike (spacelike).

\paragraph{Standard cosmologies.} We can promptly recall a couple of known examples.  
In the de Sitter case $H(t)=H_0$ and constant, so that $d\gamma_H^2$ vanishes identically: hence its null character.  

In the one-fluid model, one has 
\beq
H^2=\frac{\chi}{3} \rho=H^2_0 e^{-3(1+\omega)y}\,, \quad H^2_0=\frac{\chi c}{3}\,.
\eeq
Thus
\beq
R_H(y)=\frac{1}{H_0}e ^{\frac{3(1+\omega)}{2}y}\,,
\eeq
and the line element
\begin{equation*}
d\gamma_H^2 = \frac{9}{4 }R^2_H\, (1+\omega ) \left(-\frac{1}{3}+\omega \right)\, dy^2 \, ,
\end{equation*}
so that the horizon is timelike for $-1<\omega<1/3$.  The values $\omega=-1$ (cosmological constant) and $\omega=1/3$ (radiation-dominated cosmologies) give the horizon a null character.  On the other hand, models with $\omega>1/3$ (including stiff matter) contain a spacelike horizon.

\paragraph{Big Bang and/or Big Rip.}  Here we consider the model containing Big Bang as well as Big Rip solutions that has been presented in the previous section:  the three-fluids model given in eq.(\ref{doublefluid}).  

We recall that
\beq
R_H(y)=\sqrt{\frac{\chi}{3}}\frac{e^{3y/2}}{D^{1/2}}\,,\quad D=c_0+c_re^{-y}+ c_fe^{(3+3\delta)y}\,.
\eeq
Thus
\beq
\frac{dR_H}{dy}=\frac{R_H}{2D} N\,, \quad N=\left(3D-\frac{d D}{dy} \right)\,,
\eeq
and one has
\beq
d\gamma_H^2=-\frac{ R^2_H(y)}{4D^2}\left(c_0+(4+3\delta)c_fe^{(3+3\delta)y}  \right)N\,dy^2\,. 
\eeq
As a consequence the causal nature of the horizon depends on $N$.
For $\omega_f=-1$ or $\delta=0$ (the standard $\Lambda$CDM model) it turns out that
\beq
N=3c_0+4c_re^{-y} >0\,.
\eeq
Thus, in this case the Hubble horizon is timelike, approaching a null character (\textit{i.e.} $d\gamma_H^2\rightarrow0$) for $y\rightarrow\infty$. The same fact holds true for the Chaplygin gas model.  This is consistent with the results of the previous section.  In the phantom scenario, one has 
\beq
N=3c_0+4c_re^{-y}-3\delta c_fe^{(3+3\delta)y}\,.
\eeq
A direct calculation, making  use of equation (\ref{max}) leads to
\beq
N=-\frac{3c_0}{e^{(3+3\delta)y^*}}\left( e^{(3+3\delta)y}- e^{(3+3\delta)y^*}  \right)-\frac{4c_r e^{-y}}{e^{(4+3\delta)y^*}} \left( e^{(4+3\delta)y}- e^{(4+3\delta)y^*}  \right) \,. 
\eeq
Thus, if $y <y^*$ the Hubble horizon is timelike, if $y=y^*$ it is null and for $y >y^*$ it is spacelike.  These three ranges correspond to the ranges in which the horizon's radius is increasing, instantaneously stationary and decreasing, respectively.  Again, given that the light rays can cross the horizon toward the origin only if the horizon itself is timelike and increasing, the range in which this can occur is $y <y^*$, the same result of the previous section.

\paragraph{Time dependent dark energy model.}

Eventually, it may be of some interest to discuss the example presented in \cite{lewis1}.  While it is not a physically motivated scenario, it is nevertheless an example that shows the dependence on the Hubble horizon of the dynamics. The model is defined by the usual Friedmann equations and equations of state for ordinary matter and the ``dark energy'' component with a suitable time dependent barotropic factor:
\beq
H^2=\frac{\chi}{3}(\rho_M +\rho_E)\,, \quad p_M=0\,, \quad p_E=\omega(t)\rho_E\,, \quad \omega(t)=-1+\delta+g(t)\,,
\eeq   
where $ \delta >0$ but small (strictly dark energy contribution) and $g(t) >0$ (a non-dark part). Assuming energy conservation for  $\rho_M$ and $\rho_E$,
\beq
\dot{H}=-\frac{\chi}{2}\Big(\rho_M +\left(1+\omega(t)\right)\, \rho_E \Big)\,. 
\eeq   
As a result, the evolution of the Hubble horizon is given by
\beq
\dot{R}_H=\frac{3}{2}\frac{\rho_M +(-\delta+g(t))\rho_E }{(\rho_M +\rho_E)}\,.
\eeq   
The model is not exactly solvable due to the time dependence of $\omega(t)$.
Also  the use of the evolution parameter $y$, as in our approach, does
not simplify the computation,
and it seems difficult  to reproduce analytically the results of
Ref.~\cite{lewis1}.   However, as reported  in the cited paper, one may distinguish between two regimes: the first one valid for small values of $t$, when  the dark energy component can be neglected, and the second one valid for large values of $t$, when the dark energy component dominates. 

In the first regime one obtains $R_H\simeq \frac{3}{2}t$, while in the
second regime one has
\beq
R_H \simeq \frac{3}{2}\int_{t_1}^t (-\delta+g(t'))dt'\,,
\eeq
valid for $t_1$ sufficiently large. If the positive function $g(t)$ is
locally summable then $R_H$ may eventually diverge at infinity only. Moreover,
if $g(t_2)=\delta$ for $t_2\in(t_1,\infty)$, then
$R_H(t_2)$ is  a local minimum for $R_H$ and in such a case the behavior of $R_H$ in
the whole range $t\in(0,\infty)$ is the one considered in \cite{lewis1}.
In that paper the function $g(t)$ has been chosen in such a way that  
for very large $t$ the dark energy contribution becomes negligible and the universe falls down 
in  the standard exapanding phase with non a negative barotropic parameter. 
Thus there exist photon trajectories which cross the dynamical horizon
and arrive at the origin. 
As we already said in the Introduction this is not a surprising behavior, because the causal nature of Hubble surface may change according to Eq. (\ref{coglio}) or (\ref{gh}).

\section{Conclusions}
\label{conc}

We start our concluding remarks by discussing one of the perhaps confusing concepts that has risen in the debate, that is the \textit{present size of the horizon}.  It should be clear that if one takes into account limits of observation for an observer sitting in the origin at time $t_0$, the size of the horizon at that time, \textit{i.e.} $R_H(t_0)$, plays no role: the information about what happens at $R_H(t_0)$ will reach the origin at a time $t>t_0$ (if the horizon allows it, and this depends on the future behavior of the model).

In this paper we focused on the analysis of particular physically motivated models presenting a Big Bang singularity and different future behavior, a Big Rip ($\delta>0$ three-fluid model) or no singularity ($\delta=0$ and generalized Chaplygin gas models): in any case we restricted the analysis to expanding universes, clearly the most interesting in view of the present behavior of our own Universe.  With these models in mind, we showed that the present-day observational horizon (identified by the maximum proper radius attained by the ingoing light paths reaching the origin now) cannot be larger than the maximum proper radius attained by the Hubble horizon, which is what eq.(\ref{bound}) expresses.  One has to keep in mind that light rays are able to cross the horizon toward the origin only if the horizon is increasing (and timelike). With regard to this issue, we have confirmed the results  presented in \cite{lewis1}.  Furthermore, we have provided a sufficient condition for incoming light rays to reach the origin through the general condition $C<C^*$ [see Eq.(\ref{C}) and discussions], which actually applies to every expanding model with a Big Bang initial singularity, including also  the old one-fluid standard model, where  $C^*=\infty$, and for which the condition is always satisfied.

\end{document}